# Room-temperature intrinsic nonlinear planar Hall effect in TaIrTe$_4$

Chang Jiang[1,*], Fan Yang[2,*], Jinshan Yang[3,*], Peng Yu[4], Huiying Liu[2], Yuda Zhang[1], Zehao Jia[1], Xiangyu Cao[1], Jingyi Yan[3], Zheng Liu[5], Xian-Lei Sheng[2,6,†], Cong Xiao[7,‡], Shengyuan A. Yang[8], Shaoming Dong[3], and Faxian Xiu[1,9,10,11,#]

[1]State Key Laboratory of Surface Physics and Department of Physics, Fudan University, Shanghai 200433, China
[2]School of Physics, Beihang University, Beijing 100191, China
[3]State Key Laboratory of High Performance Ceramics and Superfine Microstructure, Shanghai Institute of Ceramics, Chinese Academy of Science, Shanghai 200050, China
[4]School of Materials Science and Engineering, Sun Yat-sen University, Guangzhou 510275, China
[5]School of Materials Science and Engineering, Nanyang Technological University, Singapore, Singapore
[6]Peng Huanwu Collaborative Center for Research and Education, Beihang University, Beijing 100191, China
[7]Interdisciplinary Center for Theoretical Physics and Information Sciences (ICTPIS), Fudan University, Shanghai 200433, China
[8]Research Laboratory for Quantum Materials, Department of Applied Physics, The Hong Kong Polytechnic University, Hong Kong, China
[9]Institute for Nanoelectronic Devices and Quantum Computing, Fudan University, Shanghai 200433, China
[10]Zhangjiang Fudan International Innovation Center, Fudan University, Shanghai 201210, China
[11]Shanghai Research Center for Quantum Sciences, Shanghai 201315, China

*These authors contributed equally to this work.
†Contact author: xlsheng@buaa.edu.cn
‡Contact author: congxiao@fudan.edu.cn
#Contact author: faxian@fudan.edu.cn

**ABSTRACT**. Intrinsic responses are of paramount importance in physics research, as they represent the inherent properties of materials, independent of extrinsic factors that vary from sample to sample, and often reveal the intriguing quantum geometry of the band structure. Here, we report the experimental discovery of a new intrinsic response in charge transport, specifically the intrinsic nonlinear planar Hall effect (NPHE), in the topological semimetal TaIrTe$_4$. This effect is characterized by an induced Hall current that is quadratic in the driving electric field and linear in the in-plane magnetic field. The response coefficient is determined by the susceptibility tensor of Berry-connection polarizability dipole, which is an intrinsic band geometric quantity. Remarkably, the signal persists up to room temperature. Our theoretical calculations show excellent agreement with the experimental results and further elucidate the significance of a previously unknown orbital mechanism in intrinsic NPHE. This finding not only establishes a novel intrinsic material property but also opens a new route toward innovative nonlinear devices capable of operating at room temperature.

Intrinsic responses have always been a focus of research in condensed matter physics. A prominent example is the intrinsic anomalous Hall effect in magnetic materials, which manifests the Berry curvature of the band structure [1]. Recently, it has been discovered that in magnetic materials with broken inversion symmetry, there also exists an intrinsic nonlinear Hall effect, where the Hall current is quadratic in the driving electric field [2], *i.e.*, $j_H \sim E^2$. The intrinsic response coefficient reflects the dipole moment of Berry-connection polarizability [3,4] (BCP), which is an important band geometric quantity underlying a range of nonlinear responses [5–9]. This effect has been successfully detected in several magnetic systems, such as few-layer antiferromagnet MnBi$_2$Te$_4$ [10,11] and noncollinear antiferromagnet Mn$_3$Sn [12].

In nonmagnetic materials, the intrinsic nonlinear Hall effect is forbidden by the time-reversal symmetry. This raises a question: What types of intrinsic nonlinear responses can be observed in such systems? One candidate is the nonlinear planar Hall effect [13] (NPHE). In the planar Hall setup, the applied magnetic field is within the transport plane, so the ordinary Hall effect from the Lorentz force is not effective here. For NPHE, the induced Hall current scales as $j_H \sim E^2 B$. From symmetry analysis, a recent theory [14] predicted that intrinsic NPHE can exist in a wide range of nonmagnetic crystals, which is much less constrained than the intrinsic nonlinear Hall effect. Previous experiments have reported NPHE measurements in systems such as Bi$_2$Se$_3$ [13,15], the two-dimensional electron gas on the SrTiO$_3$(001) surface [13], WTe$_2$ [13,16], SrIrO$_3$ [17,18], and Te [19,20]. However, the signals detected to date are of an extrinsic nature [21], most showing a scaling with relaxation time as $\sim \tau^2$; and till now, the intrinsic NPHE has not been observed yet.

To elucidate the origin of intrinsic NPHE, we connect it to the intrinsic nonlinear Hall effect. As previously mentioned, the latter is attributed to the BCP dipole. Namely, for a Hall current in $y$ direction driven by $E$ field in $x$ direction, we have $j_y^{int} = \mathfrak{D}_{yxx} E_x E_x$, where the BCP dipole is defined as [4] (take $\hbar = e = 1$)

$$\mathfrak{D}_{yxx} = \int [d\mathbf{k}] f_0'(v_y G_{xx} - v_x G_{yx}). \quad (1)$$

Here, the integration notation denotes summation over crystal momentum **k** and band index, $f_0'$ is the energy derivative of the Fermi-Dirac distribution function, $G_{ab} = 2\text{Re} \sum_{m \neq n} v_a^{nm} v_b^{mn}/(\varepsilon_n - \varepsilon_m)^3$ is the BCP tensor for band $n$ ($a, b$ are Cartesian indices), $v_a^{nm}$ is the velocity matrix element, and $\varepsilon_n$ is the band energy. Like the Berry curvature, BCP reflects interband coherence. Hence, both BCP and its dipole are expected to be enhanced in small-gap regions of the band structure. For example, in Figs. 1(b)-1(d), we illustrate the $k$-resolved BCP dipole $\mathfrak{D}_{yxx}(\mathbf{k})$, *i.e.*, the integrand in Eq. (1), for a two-dimensional gapped Dirac model (Fig. 1(a)). The results clearly indicate significant values around the small-gap region.

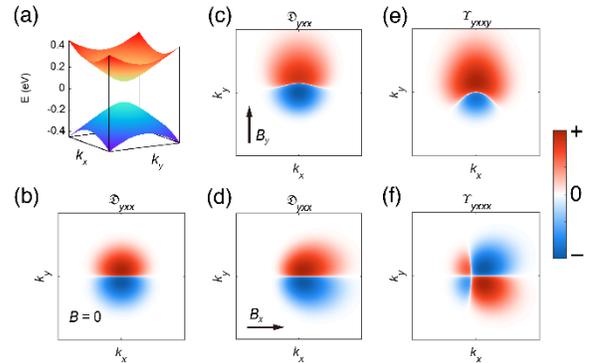

FIG. 1. Illustration of the BCP dipole triggered by an in-plane magnetic field. The effect of in-plane magnetic field on the $k$-resolved BCP dipole in a massive Dirac model, described by the Hamiltonian $H = tk_x\sigma_x + tk_y\sigma_y + \Delta\sigma_z$, is presented, where t represents the characteristic velocity and $\Delta = 0.1\text{eV}$ denotes the energy gap. (a) Band structures of the model. (b) k-resolved BCP dipole $\mathfrak{D}_{yxx}$ of the occupied band without a magnetic field. (c) and (d) k-resolved BCP dipole $\mathfrak{D}_{yxx}$ with a magnetic field applied along (c) the $y$ direction and (d) the $x$ direction. (e) and (f) k-resolved BCP dipole susceptibilities (e) $\Upsilon_{yxxy}$ and (f) $\Upsilon_{yxxx}$.

*These authors contributed equally to this work.
†Contact author: xlsheng@buaa.edu.cn
‡Contact author: congxiao@fudan.edu.cn
#Contact author: faxian@fudan.edu.cn

However, the observed pattern is antisymmetric along the $y$-direction, as shown in Figs. 1(b) and 1(d), leading to the cancellation of the obtained BCP dipoles upon integration for this model. The vanishing BCP dipole is a result of time-reversal symmetry. Actually, the BCP dipole requires breaking both inversion and time reversal symmetries. In a nonmagnetic and non-centrosymmetric crystal, one can induce a finite BCP dipole by applying a magnetic field. Indeed, as shown in Fig. 1(c), an applied in-plane magnetic field along the $y$ direction can distort the pattern and induce a nonzero BCP dipole. At low fields, the induced BCP dipole scales linearly with $B$. We can express this relation as

$$\mathfrak{D}_{abc} = Y_{abcd} B_d, \tag{2}$$

where repeated indices are summed over. Consequently, we derive the intrinsic NPHE current $j_a^{int} = Y_{abcd} E_b E_c B_d$. The fourth-rank tensor $Y_{abcd}$ is the BCP dipole susceptibility, which is also an intrinsic band geometric property (see Supplemental Material, Note 1 [31] for a detailed expression). In Figs. 1(e) and (f), we present the $k$-space patterns of the tensor components $Y_{yxxy}$ and $Y_{yxxx}$, which are concentrated in the small-gap region. Notably, $Y_{abcd}$ captures not only the Zeeman coupling between the $B$ field and the spin moment but also the coupling with the orbital moment of Bloch electrons. In the case of TaIrTe$_4$, we will demonstrate that the orbital mechanism plays a crucial role in the intrinsic response.

Symmetry analysis shows that all polar and chiral crystal classes support the intrinsic NPHE [14]. Here, we choose TaIrTe$_4$, a polar topological semimetal [22–24], which is expected to be a good platform for enhancing the BCP dipole susceptibility and generating a large NPHE. TaIrTe$_4$ is an air-stable layered material [25,26] with a $T_d$ structure (Fig. 2(a)). Its bulk crystal belongs to the space group $Pmn2_1$ and exhibits a mirror $M_a$ and a glide mirror $\widetilde{M}_b$. We take $x(y)$ direction to align with the crystal $b(a)$-axis in below.

TaIrTe$_4$ samples with thicknesses ranging from 30 nm to 40 nm (Supplemental Material, Note 2 [31]) were fabricated into Hall bar and circular disk geometries (Fig. 2(b)), with electrodes aligned along crystal axes (see Supplemental Material, Note 1 [31] for details of the sample growth and device fabrication). To measure the NPHE signals, we first examined a Hall bar device. A harmonic current was applied at a fixed frequency of 37.171 Hz along the $y$ direction, while the second-harmonic voltage drop was measured along the $x$ direction at a temperature of 300 K (see

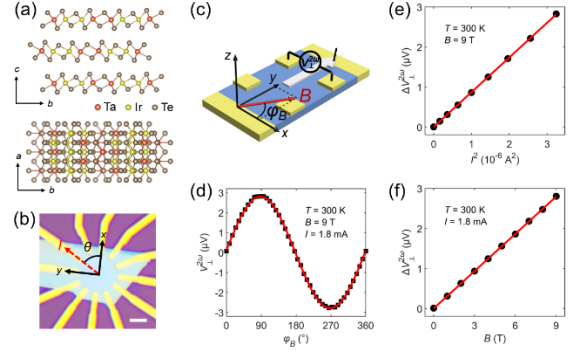

FIG. 2. Crystal structure, device configuration, and field-induced NPHE. (a) Schematic representation of the $cb$- and $ab$-plane of few-layer TaIrTe$_4$. (b) Optical image of a typical circular disk device used in this study. The scale bar is 5 μm. The current direction is indicated by the angle $\theta$ relative to the $x$ direction. The $x(y)$ direction is taken to align with the crystal $b(a)$-axis. (c) The experimental configuration of a Hall-bar device. (d) A sinusoidal dependence of $V_\perp^{2\omega}$ on the angle $\varphi_B$ when rotating the magnetic field $B$ in the $xy$-plane. The background, independent of the magnetic field, is subtracted for clarity. The symbols are the experimental data and the red lines are sinusoidal fits. (e) and (f) A linear dependence of the amplitude $\Delta V_\perp^{2\omega}$ on (e) the square of the current and (f) the magnetic field. The symbols are the experimental data and the red lines are linear fits. The uncertainties are smaller than the size of the symbols.

Supplemental Material, Note 1 [31] for details of the transport measurements). When the magnetic field is rotated in the transport plane, characterized by the angle $\varphi_B$ defined from the $x$ direction (as shown in Fig. 2(c)), the second-harmonic Hall voltage $V_\perp^{2\omega}$ exhibits a form of a sine function (Fig. 2(d)). The results were found to be independent of the driving frequency (Supplemental Material, Note 3 [31]), which excludes possible capacitive effects [5,27]. Notably, the sign and magnitude of $V_\perp^{2\omega}$ can be feasibly adjusted by rotating the magnetic field. Specifically, $V_\perp^{2\omega}$ reaches its maximum (minimum) when the magnetic field is parallel (antiparallel) to the current, and vanishes when the field is perpendicular to the current. We rule out the contribution from the out-of-plane component of the magnetic field by rotating the field in the $zy$- and $zx$-planes (Supplemental Material, Note 4 [31]). Besides, we observed a background signal in $V_\perp^{2\omega}$ that is independent of the magnetic field (Supplemental Material, Note 5 [31]). This background can be attributed to the nonsymmorphic symmetry ($\widetilde{M}_b$) breaking at the surfaces of TaIrTe$_4$, consistent with previous reports [28]. Below, we focus on the second-harmonic Hall signal induced by the magnetic field.

To reveal characteristics of the field-dependent second-harmonic Hall voltage $V_\perp^{2\omega}$, we measured it as a function of the magnitudes of both the magnetic field

*These authors contributed equally to this work.
†Contact author: xlsheng@buaa.edu.cn
‡Contact author: congxiao@fudan.edu.cn
#Contact author: faxian@fudan.edu.cn

$B$ and the driving current $I$. By fitting $V_\perp^{2\omega}$ as a function of the angle $\varphi_B$ with the sinusoidal function, we extracted its amplitude $\Delta V_\perp^{2\omega}$. At a fixed field $B = 9$ T, $\Delta V_\perp^{2\omega}$ increases linearly with the square of the driving current, consistent with its electrically nonlinear origin, as shown in Fig. 2(e). At a fixed current $I = 1.8\ mA$, $\Delta V_\perp^{2\omega}$ also increases linearly with the magnetic field (Fig. 2(f)), and this linearity persists up to at least 9 T across all samples we measured. This $j_H \sim E^2 B$ scaling is consistent with the NPHE we aim to investigate.

Next, to investigate the angular dependence of the NPHE, we adopted a circular disk geometry and conducted the measurement at 300 K. The driving current was applied between two opposite electrodes, with the current direction defined by angle $\theta$ measured from the $x$ direction (Fig. 2(b)). Voltage signals at electrodes parallel and perpendicular to the driving current were measured at both first- and second-harmonic frequencies. We show the results from disk device S3 in Fig. 3. First, we determine the crystal axes and examine the anisotropy of resistance. From the first-harmonic signals, we derive the longitudinal resistance $R_\parallel \equiv V_\parallel/I$ and the transverse resistance $R_\perp \equiv V_\perp/I$ (Fig. 3(a)). $R_\perp$ can be nonzero due to the misalignment of the electrodes with the crystal axes and the intrinsic anisotropy of the material (Supplemental Material, Note 6 [31]). Both the longitudinal and transverse resistances exhibit a two-fold angular dependence. This observation aligns with the $C_{2v}$ symmetry of TaIrTe$_4$, which gives $R_\parallel(\theta) = R_y \sin^2\theta + R_x \cos^2\theta$ and $R_\perp(\theta) = (R_y - R_x)\sin\theta\cos\theta$, where $R_y$ and $R_x$ ($> R_y$) are the resistances along the $y$ and $x$ directions. The resistance anisotropic ratio $r (\equiv R_y/R_x)$ is determined to be about 0.24 from fitting. Next, we analyze the second-harmonic voltage $V_\perp^{2\omega}$, which maintains a sinusoidal dependence on $\varphi_B$ at all current orientations, as shown in Fig. 3(b). A noticeable feature is that when the current is applied along the $y$ (or $x$) direction, $V_\perp^{2\omega}$ shows a sine (or cosine) dependence on the direction of the magnetic field. This observation again complies with the $C_{2v}$ symmetry of TaIrTe$_4$, which constrains the angular dependence of $V_\perp^{2\omega}$ as (see Supplemental Material, Note 1 and Note 7 [31] for derivation),

$$\frac{V_\perp^{2\omega}}{(V_\parallel)^2 B} \propto \frac{-r\chi_1 \sin\theta \sin\varphi_B + \chi_2 \cos\theta \cos\varphi_B}{(\cos^2\theta + r\sin^2\theta)^2}, \quad (3)$$

where $\chi_1 \equiv \chi_{xyyy}^{(2)}$ and $\chi_2 \equiv \chi_{yxxx}^{(2)}$ are the two independent symmetry-allowed elements of NPHE conductivity tensor, defined by $j_a^{(2)} = \chi_{abcd}^{(2)} E_b E_c B_d$. By rewriting the right side of Equation (3) in the form of $A\cos(\varphi_B - \beta)$, we can separately analyze the angle $\beta$ and the amplitude $A$ as a function of the current orientation $\theta$ (Supplemental Material, Note 8 [31]). The experimental data for $\beta$ and $A$ fit well with Equation (3) and align with the crystal axis directions determined by first-harmonic measurement, as shown in Figs. 3(c) and 3(d). These agreements further validate our measurements and analysis. In Fig. 3(c), rotating the current direction reveals $\beta \neq \theta$, consistent with Equation (3) when $|r\chi_1| \neq |\chi_2|$. While misalignment of electrodes could lead to similar behavior, we have ruled out this possibility through temperature-dependent measurements (Supplemental Material, Note 9 [31]). The amplitude $A$ reaches its maximum (or minimum) when the current is aligned with $y$ (or $x$) direction, as shown in Fig. 3(d), implying that $r\chi_1$ is greater than $\chi_2$. Consistent results were observed across all disk devices we measured (Supplemental Material, Note 12 [31]). It is noteworthy that previously reported extrinsic NPHE in Bi$_2$Se$_3$ [13] and SrIrO$_3$ thin films [17] exhibited nonlinear resistance $R_{yx}^{2\omega} \sim \mathbf{E} \cdot \mathbf{B}$ with a different angular dependence. The distinct feature $\beta \neq \theta$ here reflects different crystal symmetries and also suggests a possibly different microscopic origin.

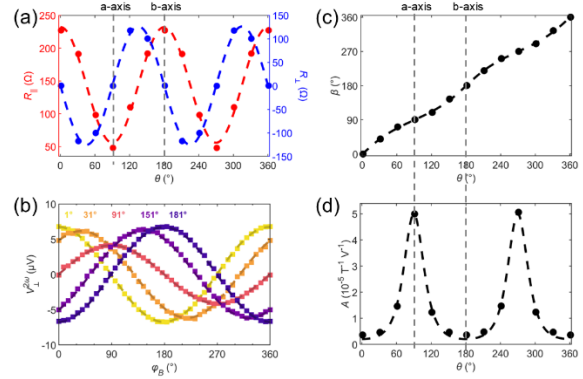

FIG. 3. Angular dependence of the NPHE at 300 K. (a) $R_\perp$ and $R_\parallel$ of a circular disk device as a function of the current direction $\theta$. The symbols are the experimental data and the dashed lines are sinusoidal fits. (b) Sinusoidal dependence of $V_\perp^{2\omega}$ on the angle $\varphi_B$ for different current directions. The symbols are the experimental data and the dashed lines are sinusoidal fits. The background, independent of the magnetic field, is subtracted for clarity. (c) and (d) (c) The angle $\beta$ and (d) the amplitude $A$ as a function of $\theta$, derived from $V_\perp^{2\omega}/(V_\parallel^2 B) = A\cos(\varphi_B - \beta)$. The symbols are the experimental data and the dashed lines are fits with the model described in the text. The uncertainties are smaller than the size of the symbols.

The observed NPHE in TaIrTe$_4$ is pronounced and persists at room temperature. The NPHE conductivity $\chi_1$ reaches approximately $10^{-4} A \cdot T^{-1} \cdot V^{-2}$, which is two orders of magnitude larger than the extrinsic

*These authors contributed equally to this work.
†Contact author: xlsheng@buaa.edu.cn
‡Contact author: congxiao@fudan.edu.cn
#Contact author: faxian@fudan.edu.cn

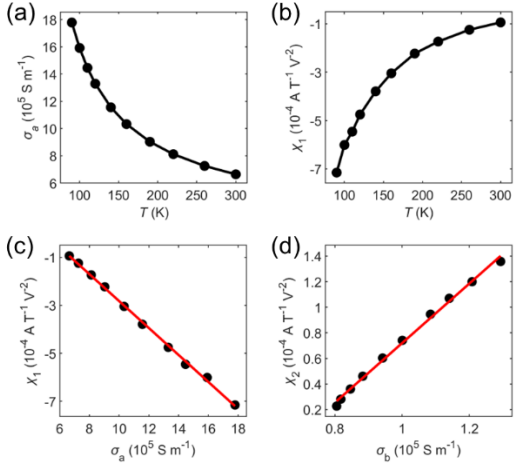

FIG. 4. Temperature-dependence measurement and scaling of NPHE. (a) and (b) Temperature dependence of (a) $\sigma_a$ and (b) $\chi_1$. The symbols are experimental data and the lines are guides to the eye. (c) and (d) The nonlinear planar Hall conductivity (c) $\chi_1$ as a function of $\sigma_a$ and (d) $\chi_2$ as a function of $\sigma_b$. The symbols are experimental data and the red lines are linear fits. The uncertainties are smaller than the size of the symbols.

NPHE reported in SrIrO$_3$ thin films [17]. Such a large effect here is unlikely due to the extrinsic Drude-like contribution, which tends to be suppressed at room temperature, as previously demonstrated in Bi$_2$Se$_3$ [13] and SrIrO$_3$ thin films [17]. Indeed, our first-principles calculations (Supplemental Material, Note 1 [31]) estimate that this extrinsic Drude-like contribution is two orders of magnitude smaller than the experimental NPHE. On the other hand, due to the topological semimetal character of TaIrTe$_4$, with many small-gap and band-crossing regions around the Fermi level (Supplemental Material, Note 13 [31]), it is reasonable to expect that the intrinsic mechanism arising from enhanced band geometric quantities may play a significant role in the observed behavior.

To further investigate the possible mechanisms underlying the observed NPHE, we examine the scaling behavior of the NPHE conductivity $\chi$ with respect to the longitudinal conductivity $\sigma$. As $\sigma$ shows no obvious dependence on gate voltage (Supplemental Material, Note 14 [31]), we vary $\sigma$ by changing the temperature, from 300 K to 100 K. The results from disk device S4 are presented in Fig. 4. $V_\perp^{2\omega}$ maintains sinusoidal dependence on the angle $\varphi_B$ at all temperatures (Supplemental Material, Note 9 [31]). Since only $\chi_1$ (or $\chi_2$) contributes to $V_\perp^{2\omega}/(V_\parallel^2 B)$ when the driving current is along $y$ (or $x$) direction, we compare the temperature dependences of $\chi_1$ and $\sigma$ along the $y$ direction, namely $\sigma_a$ of the crystal (Figs. 4(a) and 4(b)). The absolute values of both show a monotonic decrease with temperature. As shown in

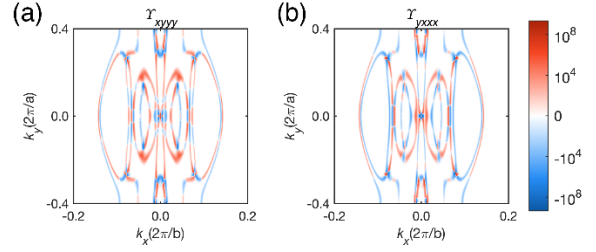

FIG. 5. First-principles results for in-plane susceptibility of the BCP dipole in TaIrTe$_4$. Calculated $k$-resolved contributions to the BCP dipole susceptibility (Eq. (10) in Methods, in units of $10^{-9} \frac{e^2}{\hbar} \cdot Å^3 \cdot \Omega^{-1} \cdot T^{-1} \cdot V^{-1}$ ). (a) $\Upsilon_{xyyy}$ and (b) $\Upsilon_{yxxx}$ on the Fermi surface in the $k_z = 0$ plane of the Brillouin zone. The prominent distributions are contributed by band near-degenerate regions on the Fermi surface. Only $\Upsilon_{xyyy}$ and $\Upsilon_{yxxx}$ contribute to transport, reflecting the control of the NPHE by the magnetic field orientation.

Figs. 4(c) and 4(d), we observe an almost perfect linear scaling relation between the two:

$$\chi = \xi\sigma + \eta, \qquad (4)$$

where $\xi$ and $\eta$ are scaling parameters. Such a scaling behavior has not been reported for the NPHE previously, contrasting with the quadratic scaling attributed to the nonlinear Drude contribution in prior reports [13,16,20].

Our primary interest lies in the zeroth-order term $\eta$ in the scaling relation, which captures the intrinsic contribution $\Upsilon$. Fitting the data in Figs. 4(c) and 4(d) yields $\eta_1 = 2.8 \times 10^{-4} \, A \cdot T^{-1} \cdot V^{-2}$ and $\eta_2 = -1.6 \times 10^{-4} \, A \cdot T^{-1} \cdot V^{-2}$, corresponding to $\chi_1$ and $\chi_2$, respectively. This analysis was conducted for all three circular devices we fabricated, and the extracted $\eta$ values are of similar magnitudes (Supplemental Material, Table S1 [31]). Meanwhile, we performed first-principles calculations to evaluate the intrinsic contribution in TaIrTe$_4$ (Supplemental Material, Note 1 [31]). In Fig. 5, we present the result of $k$-space distribution for the relevant $\Upsilon$ tensor components. Our findings indicate that the large contributions arise from small-gap regions on the Fermi surface. After the $k$-space integral, we obtain $\chi_1^{int} \equiv \Upsilon_{xyyy} = 2.40 \times 10^{-4} \, A \cdot T^{-1} \cdot V^{-2}$ and $\chi_2^{int} \equiv \Upsilon_{yxxx} = -1.73 \times 10^{-4} \, A \cdot T^{-1} \cdot V^{-2}$. The agreement between theory and experiment is excellent: not just the magnitudes match closely, the theory also captures the correct signs of the two conductivity components. This provides strong evidence that the contribution $\eta$ here is dominated by the intrinsic NPHE response we are looking for. This NPHE in TaIrTe$_4$ enables us, for the first time, to probe the BCP dipole susceptibility $\Upsilon$.

*These authors contributed equally to this work.
†Contact author: xlsheng@buaa.edu.cn
‡Contact author: congxiao@fudan.edu.cn
#Contact author: faxian@fudan.edu.cn

Importantly, our calculations reveal that the coupling between the magnetic field and the orbital moment significantly contributes to the intrinsic NPHE in TaIrTe$_4$, compared to the spin contribution from Zeeman coupling. For instance, in $Y_{yxxx}$, the orbital contribution is $-1.46 \times 10^{-4} A \cdot T^{-1} \cdot V^{-2}$, which is over one order of magnitude larger than the spin contribution. This orbital mechanism has not been reported previously in the NPHE, which has generally been attributed solely to the spin Zeeman coupling. Notably, distinct from the spin mechanism, the orbital mechanism identified here does not require spin-orbit coupling. This implies that it may lead to pronounced NPHE even in materials with negligible spin-orbit coupling strength.

Finally, besides the intrinsic NPHE, our experimental results (Fig. 4(c) and 4(d)) also show an extrinsic contribution exhibiting linear scaling with $\sigma$. Such a contribution is generally attributed to non-Gaussian conventional skew scattering when static disorder (impurities, crystal imperfections, boundary roughness, etc.) is the dominating scattering source. However, this scenario seems unlikely here, given the high quality of our samples and the persistence of the linear behavior up to room temperature, where phonon scattering is expected to dominate. An alternative explanation could involve competition between two or more scattering processes, such as electron-impurity and electron-phonon scattering. Recent theory [29] suggests that the compositions of side jump or intrinsic skew scattering with Berry curvature anomalous velocity, as well as their electric field corrections, may also lead to a linear scaling behavior. Further efforts will be necessary to elucidate the microscopic origin of this linear extrinsic contribution in future studies.

In conclusion, we have demonstrated a new intrinsic transport response, the intrinsic NPHE, in the topological semimetal TaIrTe$_4$. This discovery allows us to probe the BCP dipole susceptibility, an intrinsic band structure property that encodes the quantum geometry of Bloch electrons. Furthermore, we unveil a novel orbital mechanism that operates independently of spin-orbit coupling. Compared to the extrinsic nonlinear Hall effect arising from Berry curvature dipole and the intrinsic nonlinear Hall effect, intrinsic NPHE may be realized in a much broader class of materials, providing a new tool for characterizing their intrinsic properties. This is particularly helpful for studying emerging polar and chiral topological semimetals, which have gained significant theoretical and experimental interests. Many of these materials exhibit relatively high symmetries, such as $C_{6v}$, $D_6$, $T$ and $O$ point groups, that forbid electrical second- and third-order nonlinear Hall effects but allow for NPHE. Moreover, this effect also opens a new pathway for the development of room-temperature nonlinear devices, which may find promising applications in wireless rectification and energy harvesting [28,30].

ACKNOWLEDGMENTS

F. X. was supported by the National Natural Science Foundation of China (52225207 and 52350001), the Shanghai Pilot Program for Basic Research - FuDan University 21TQ1400100 (21TQ006), and the Shanghai Municipal Science and Technology Major Project (Grant No.2019SHZDZX01). J. Y. was supported by the National Natural Science Foundation of China (Grant No. 52222202). H.L. acknowledges support from the National Natural Science Foundation of China (Grant No. 12304053). Z.L. acknowledges support from the Singapore Ministry of Education Tier 3 Program "Geometrical Quantum Materials" AcRF Tier 3 (MOE2018-T3-1-002). X.-L.S. acknowledges support from National Key R&D Program of China (Grant No. 2022YFA1402600), and the NSFC (Grants No. 12174018). C.X. acknowledges support from the start-up funding from Fudan University. S.A.Y acknowledges support from Macau FDCT Funding Scheme for Scientific Research and Innovation (0066/2024/RIA1). We thank Dong Sun from Peking University for helpful discussions.

*These authors contributed equally to this work.
†Contact author: xlsheng@buaa.edu.cn
‡Contact author: congxiao@fudan.edu.cn
#Contact author: faxian@fudan.edu.cn

*These authors contributed equally to this work.
†Contact author: xlsheng@buaa.edu.cn
‡Contact author: congxiao@fudan.edu.cn
#Contact author: faxian@fudan.edu.cn

angular dependence of $V_\perp^{2\omega}$, fitting of the nonlinear planar Hall effect, temperature dependence of the nonlinear planar Hall effect, analysis of other nonlinear effects, additional Hall bar devices, additional circular disk devices, band structures and Fermi surface of TaIrTe4, and gate dependence of the nonlinear planar Hall effect, which includes Refs. [32-45].

*These authors contributed equally to this work.
†Contact author: xlsheng@buaa.edu.cn
‡Contact author: congxiao@fudan.edu.cn
#Contact author: faxian@fudan.edu.cn